# The Essence Theory of Software Engineering – Large-Scale Classroom Experiences from 450+ Software Engineering BSc Students

Kai-Kristian Kemell[1][0000-0002-0225-4560] Anh Nguyen-Duc[2][0000-0002-7063-9200] Xiaofeng Wang[3][N/A] Juhani Risku[1][N/A] and Pekka Abrahamsson[1][0000-0002-4360-2226]

[1] University of Jyväskylä, 40014 Jyväskylä, Finland
{kai-kristian.o.kemell|pekka.abrahamsson|juhani.risku}@jyu.fi
[2] University of Southeast Norway
angu@usn.no
[3] Free University of Bozen-Bolzano, 39100 Bozen-Bolzano, Italy,
xiaofeng.wang@unibz.it

**Abstract.** *Software Engineering as an industry is highly diverse in terms of development methods and practices. Practitioners employ a myriad of methods and tend to further tailor them by e.g. omitting some practices or rules. This diversity in development methods poses a challenge for software engineering education, creating a gap between education and industry. General theories such as the Essence Theory of Software Engineering can help bridge this gap by presenting software engineering students with higher-level frameworks upon which to build an understanding of software engineering methods and practical project work. In this paper, we study Essence in an educational setting to evaluate its usefulness for software engineering students while also investigating barriers to its adoption in this context. To this end, we observe 102 student teams utilize Essence in practical software engineering projects during a semester long, project-based course.*

## 1 Introduction

Software Engineering (SE) work out in the field is diverse, with practitioners employing a myriad of different methods and practices in equally diverse SE endeavors [5, 10]. As little consensus exists in terms of best practices and methods, practitioners have taken to using what they consider to be the best option(s) for their own SE context, often tailoring them by omitting some suggested practices or rules [5]. Though e.g. Agile methods are currently widely employed out on the field, the practices and methods that are understood as being Agile are numerous [1]. Especially software startups use a diverse mix of agile methods and practices, with some simply opting to use ad hoc SE methods [17].

This diversity in the SE industry has, alongside other factors such as technological advances, resulted in a gap between education and practice in SE [2, 13]. As it is not possible to teach university students all the methods and practices employed by practitioners, curriculum-makers are faced with choices on what to focus on. General theories and methods that can be taught to students to support them in the adoption of new practices in the future are one option in attempting to tackle this gap. One such theory is the *Essence Theory of Software Engineering* (Essence from here-on-out), proposed by the SEMAT initiative[1] [10].

Created to address the vast range of methods employed in the field, Essence is a method-agnostic progress control tool for SE. Essence is modular in nature and can be used to model any existing methods, practices, or combination of such [15]. Thus, Essence is designed to suit any SE possible context [9], making it a potentially powerful tool. However, its flexibility is also a potential a downside: in order to use Essence, resources have to be devoted towards modeling the practices and methods being used, as well as learning how to do specifically by using Essence.

Presently, Essence has yet to see widespread adoption among practitioners, although it has seen some traction among the academia [21]. It is possible that its rather resource-intensive adoption is one barrier for its adoption, as has been discussed in extant research [8, 18]. For this purpose, some tools have been suggested to aid practitioners in its adoption and in using it: e.g. [8] presented SematAcc to help users visually track the alpha states while using Essence and [11] presented an Essence-themed board game to make learning Essence easier. However, more tools and further studies specifically focusing on its supposedly difficult adoption are also required to better understand the barriers of its adoption and to consequently be able to tackle them. Additionally, an educational perspective on Essence is interesting because Essence can help address the gap between education and industry needs. For example, [2] report that SE graduates are often perceived by the industry as lacking in e.g. the ability to follow processes and project management skills, both of which Essence can help teach.

---
[1] semat.org

In this paper, we study Essence in a large-scale classroom setting. We observe over one hundred project teams consisting of second year SE students employ Essence during course projects mimicking a field SE endeavor. The teams carry out a complete SE project, from requirements formulation to a finished software product, using Essence to manage their project. Then, based on their projects, the students reflect on their experiences with Essence in a written experience report. With the data collected from these experience reports, we seek to understand:

*RQ1:* How useful do bachelor level students find Essence?

*RQ2:* What are the challenges in adopting Essence, specifically for inexperienced software developers, and what could be done to make its adoption easier?

The rest of this paper is structured as follows. In the next section, we discuss the Essence specification and extant research on it in further detail. In the third section, we present and discuss the study design. In the fourth section, we analyze the data and present our findings. We then discuss the practical and theoretical implications of our findings in the fifth section, as well as the potential limitations of the study and directions for future research. The sixth and final section concludes the paper.

## 1 The Essence Theory of Software Engineering

Essence is a modular, method-agnostic progress control tool for SE endeavors. Proposed by the SEMAT initiative to address the myriad of methods and practices employed by industry practitioners, Essence is a framework into which any combination of existing methods or practices can be inserted. In practice, Essence consists of a kernel (seen in Figure 1) and a language. The kernel [14], its authors argue [10], contains all the elements present in every SE endeavor, while the language can be used to extend the kernel to fit any specific SE endeavor. I.e. Essence, in its base form, contains the elements required to track progress in a generic SE endeavor, but it is intended to be tailored for specific SE contexts.

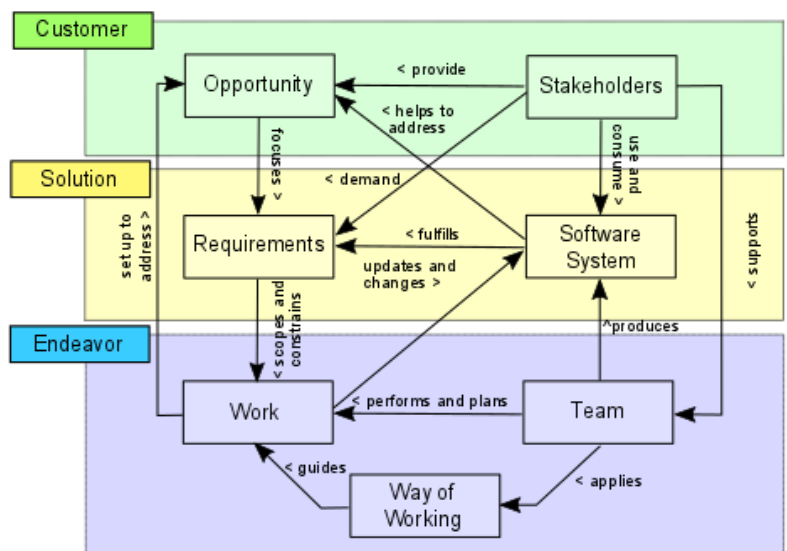

**Figure 1.** The Essence Kernel Alphas

The Essence kernel consists of three views: *alphas*, *activity spaces*, and *competencies*. In the kernel, there are seven alphas, "things to work with": opportunity, stakeholders, requirements, software system, work, team, and way of working [10]. These alphas, Jacobson et al. [10] posit, are present in every SE endeavor. Alpha is an acronym for an "Abstract-Level Progress Health Attribute" [14]. For the project to progress, these alphas need to be worked on. To this end, the kernel contains activity spaces. Activity spaces may contain 0 or n activities, or "things to do". The activity spaces in the kernel, much like the alphas, are elements Jacobson et al. [10] argue are found in every SE endeavor. Finally, the kernel contains a set of competencies: skills needed to carry out the endeavor [10]. These alphas, activity spaces, and competencies are further split into three areas of concern: *endeavor*, *solution*, and *customer*.

The alphas of the kernel serve as a way of tracking project health. *Alpha states* offer a way of tracking progress on the various areas of the endeavor. Each of the seven base alphas has a set of states that describe the progress made on each individual alpha. For example, the states for the requirements alpha range from conceived, where the requirements have only just been formulated, to fulfilled, where they have been implemented into the system in a manner satisfying the stakeholders.

Jacobson, Stimson & Hastie [9] suggest Essence as a solution to what they call method prisons. In speaking of method prisons, they refer to the idea of organizations being stuck following one method or set of methods regardless of their suitability in the current context at any given time. However, they posit, the SE practitioners often present methods as monolithic for example by using very varied presentation styles to describe them. By presenting methods in a uniform manner, by e.g. using Essence, and by simply promoting a method-agnostic idea, Jacobson et al. [9] argue that organizations could escape method prisons and potentially improve their work processes by creating better methods specifically suited for their SE context.

Though its modular and extensible nature is the greatest strength of Essence, it can also be its greatest weakness. Whereas it makes Essence a powerful tool, it also makes it both resource-intensive and potentially difficult to adopt. Perhaps consequently, Essence has not gained widespread recognition among practitioners, although it has gained some traction among the academia [21]. Graziotin & Abrahamsson [8] suggest that the modest attention Essence has received among practitioners may well stem from the steep learning curve of the specification. Even though Jacobson et al. [9] make a potentially interesting case in promoting the idea of tailoring methods more actively, it may seem easier for practitioners to get started by simply using an existing method.

## 2 Research Design and Methodology

In this section, we describe the methodology of the classroom study on Essence in the context of student SE projects. In the first sub-section, we discuss the course from which the data was collected. The role of Essence in said course is then discussed in the second sub-section. The third and final sub-section discusses our data collection and analysis methodology in detail. The data is then analyzed in the following main section.

### 2.1 The Course

The study presented in this paper was conducted using data from the *TDT4140 – Software Engineering* course at the Norwegian University of Science and Technology (NTNU). More specifically, all data for this study was collected during the 2017 spring iteration of the course during which the students utilized Essence in their projects. In this instance of the course, each project team was to engineer a functional software by carrying out a real SE project in a university setting. The theme of the projects was to radically improve university education by means of software robots. The exact goal of the projects was to "make a bot to replace Prof. Abrahamsson at his course on SE".

Following the first lecture of the course, the students were instructed to form project teams consisting of 4 to 5 students. The teams were formed by having the students give a subjective evaluation of their own programming skills in terms of programming confidence and then form teams with individuals with similar evaluations. This was done to negate any potential internal issues (e.g. workload distribution issues) within the teams arising from skill differences in programming. Starting from the first lecture, these teams were to work on their projects until the end of the course. The teams were first tasked with interviewing university teaching staff in order to discover tangible needs that could be addressed through their software. Stakeholders were involved in this fashion to make the project mimic a real SE endeavor more closely.

After gathering needs through the interviews and selecting the one(s) they wished to address, the students were to plan their development methodology and start utilizing it. During the course and the projects, weekly two-hour-lectures continued to offer relevant information and to support the project teams. The project work itself was carried out largely independently by each team.

### 2.2 The Role of Essence in the Course

Essence was introduced to the teams in the first lecture. The first lecture focused on discussing SE work in practice, specifically from the point of view of projects. During the lecture, Essence was discussed primarily in relation to its seven alphas, which were underlined to present the essential elements of an SE endeavor. In terms of methods, the students were instructed to initially work in whatever fashion they thought was best. The reasoning behind this line of action was to create fertile ground for the later adoption of Essence: by letting the teams first work in a rather unsystematic or even ad hoc fashion, they would likely be more receptive to tools that could help them systematize their way of working. I.e. having experienced unsystematic SE project work, they would better understand the need for more structured approaches to SE.

This approach, in practice, resulted in the teams largely working with various "ScrumBut"[2] approaches for the first three weeks. Their use of Scrum was likely to have stemmed from a previous course at the university having introduced them to Scrum. After three weeks of working as they saw fit without outside assistance from the teaching team, the teams were introduced to the Ivar Jacobson Practice Library[3]. They were tasked with using the practice cards (Fig. 2) from the library to re-construct their way of working and to modify it as they saw fit based on their experiences so far.

In this fashion, the teams were introduced to both the progress control aspect of Essence and its method-agnostic philosophy during the course. After the introduction of the practice cards, the use of Essence was not enforced during the project work and there were no regular check-ups to confirm its utilization. Full and correct utilization of Essence was not mandatory, and its utilization or lack thereof did not affect the grades given to the teams. All teams were instructed to utilize it to what extent they felt they could, but this was not supervised in practice. This approach was chosen to gather more unbiased data on the possible barriers of adoption in the case of Essence.

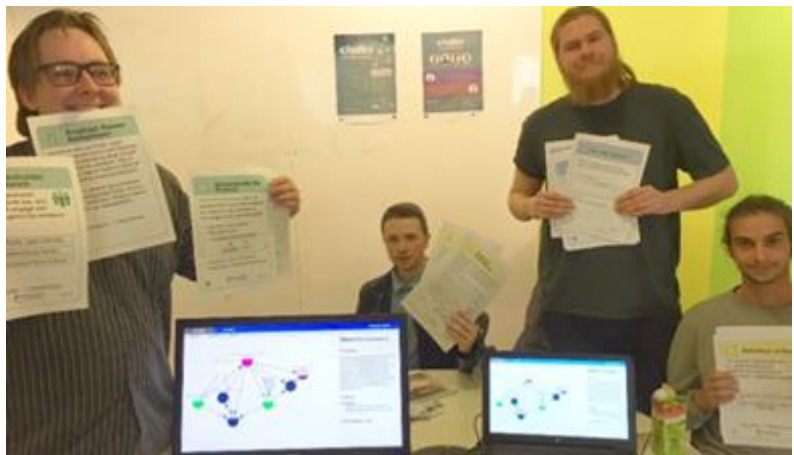

**Figure 2.** A project team showing their practice cards

### 2.3 Data Collection and Analysis Methodology

The data for this study was collected through written reflective reports provided by each team at the end of their projects[4]. In their report, each team was instructed to reflect on their experiences with Essence, along with other content unrelated to this study. As for Essence, they were to describe how they utilized it and how they felt about having done so. More specifically: (1) what they thought was good about Essence, (2) what they thought was bad about Essence, and (3) how they utilized Essence during their project.

Ultimately, 102 project teams of 4-5 students finished the course and delivered a written project report. Our data analysis is based on these 102 reports. The teams were not given a strict format to follow in the sections of their reports describing Essence, which led to the data being somewhat diverse in presentation. Each report was to discuss the afore-mentioned three topics related to their use of Essence, but past these general

---

[2] ScrumBut refers to using Scrum while omitting some parts of it, "We use Scrum, but…" (refer to: https://www.scrum.org/resources/what-scrumbut)

[3] https://practicelibrary.ivarjacobson.com/start

[4] A book showcasing the results of the projects can be found on Figshare: https://figshare.com/articles/100_Open_Sourced_Software_Robots_for_Tomorrow_s_Education_Revolutionizing_the_University_Learning_Experience_with_Bot_Technologies/5597983

guidelines the Essence sections of the reports were freeform. In practice, this largely just meant that teams that had utilized Essence relatively little wrote little about it whereas teams that had utilized it fully wrote far more about their experiences.

Thematic analysis was chosen as the method of analysis for this study due to the large volume of the data, as well as the lack of pre-determined assumptions of how the students possibly perceived the use of Essence in this context. Both the final themes and the initial codes used to formulate them were generated from the data in an inductive fashion. The analysis process was iterative and reflexive.

Initially, the author conducting the thematic analysis went through the data and recorded key points for each report, both by directly quoting the reports and by making summarizing remarks, in a separate text document. During this process, initial codes were formulated based on recurring sentiments in the reports. E.g. many reports turned out to describe various initial difficulties in adopting Essence. The analysis process was iterative, and reports and the recorded key points and quotations were regularly re-read as further codes were generated. This phase was concluded once all reports had been analyzed and the final set of codes had been applied to each of them where applicable.

Finally, the themes were generated inductively from the coded data. Codes were arranged into matching themes, with each theme encompassing one or more codes. In determining the themes, the research questions were used as a framework for organizing the data under the themes as well as determining the relevance of the codes and what was to ultimately be included into the study. In presenting the results in the next section, some of the direct quotations used in the analysis process were also included.

Additionally, in our first research question we speak of *usefulness*. Usefulness is a construct often used in relation to evaluating software systems designed especially for work-related use (e.g. [4]). In the context of this study, we define usefulness to be related to either learning something new about SE or SE progress control (educational usefulness) or providing help in SE project work (practical usefulness). These two seemingly separate types of usefulness are nonetheless closely linked together, however. E.g. a learning experience related to SE project work may simultaneously result in practical usefulness through the application the newly-learned information into practice, which may also take place at a later point in time. In our analysis, we thus speak of usefulness while referring to usefulness in both senses.

## 3   Results

The reports showed a very varying degrees and success of utilization of Essence among the 102 project teams. Whereas some of the teams had clearly utilized Essence in its entirety and reflected upon it in depth, some of the teams had done the bare minimum of selecting different practices to use while forgoing the progress control aspect of Essence. However, despite the varying degree and success of Essence utilization among the teams, the reports discussed similar themes across the spectrum.

### 3.1   Theme 1: Difficult or Resource-Intensive to Learn

The reports indicated that the majority of the teams considered Essence difficult to learn to some extent. Even most of the teams that ultimately utilized Essence successfully considered it to have been difficult to initially grasp. As the course involved only a general introduction to Essence and its principles, the teams were to study and use Essence on their own using what resources they would find on the SEMAT website or the Internet in general. This resulted in most teams feeling that Essence was difficult to learn, or "hard to get a grasp on when first introduced" (Report 048). The teams generally considered to be a direct result of the types of resources available online:

> *…we felt that almost anywhere we went to read about SEMAT we were either drowned with information (the Essence Kernel PDF has 308 pages) or the information was too abstract that we felt left confused after reading.* (Report 041)

> *The web page material, the articles and the academic resources about SEMAT are filled with many new terms, but few clear definitions. It would be easier for the next years students to grasp what SEMAT really is, if there existed some sort of document on blackboard explaining the SEMAT terminology.* (Report 016)

Largely in line with the quotation above, though Essence was considered difficult to learn, the teams almost uniformly cited the lack of good tutorial resources as the main reason for this. The existing ones were considered either too lengthy or to simply be written in a needlessly complex manner, failing to offer a good initial touch to the specification. This is also supported by some reports directly stating that past the initial barrier of adoption, Essence was a useful tool. However, due to its resource-intensive adoption, many felt that they wanted to focus on the practical SE work instead:

> *We just wanted to get on with the programming and it seemed like it was just one more unnecessary thing we needed put effort into when we already had quite a lot with learning new technologies and languages.* (Report 044)

Past the self-reported issues related to learning Essence, it was also occasionally possible to determine that a team had not managed to internalize Essence based on the contents of their report. It was evident that some teams had only utilized the practice cards, as they had been directly instructed to do, and ignored the kernel and its alphas and other views, i.e. the progress control aspect of Essence. It is likely that this was caused by the perceived difficulty of learning the specification: some of these teams likely felt that they had understood Essence despite only grasping parts of it. Though the difficulty of learning Essence was primarily blamed on the lack of good tutorial resources, one of the teams did specifically state that they felt Essence itself was too abstract for them.

Despite Essence being considered somewhat difficult to initially learn by the teams, it was generally considered to have been a positive experience. Even the teams that reported having particularly struggled with learning it, or having been unwilling to initially devote resources towards doing so, felt that it had ultimately been useful:

> *In retrospective, perhaps we would have had even greater progress with our project and higher learning outcome from the course if our understanding of SEMAT had improved at an earlier stage* (Report 062)

> *When we later, a bit too late probably, actually sat down and studied what it meant and how to use it, it seemed kind of genius.* (Report 044)

### 3.2 Theme 2: Inexperience

Another recurring theme present in the reports was inexperience in relation to SE. In their reports, the teams often discussed their own perceived inexperience with SE in relation to Essence. The inexperience of the teams evidently had a multifaceted significance to their experiences with Essence.

On one hand, the teams felt that Essence was *more* useful because they were inexperienced. They felt that, being inexperienced developers, Essence helped them (1) structure their way of working, (2) learn about new methods and practices, and (3) manage their projects better. In conjunction with the practice library, Essence was perceived to have been very educational in relation to SE methods and practices.

> *While still being on our own and with little experience, SEMAT provided us guidelines that allowed us to improve and learn while planning and working on the project. Resulting in a much better experience with projects than before and a concept we are proud of. Knowledge we absolutely will include in future projects and programming.* (Report 078)

> *...our experience with the ESSENCE kernel has been almost exclusively positive. Given that is prevents overlooking parts of the software development cycle, we perceived it as more beginner friendly than other competing, more fragmented approaches to software development methodology.* (Report 047)

On the other hand, some teams felt that their inexperience with SE might have also had a negative impact on the usefulness of Essence. As Essence encourages one to develop their own way of working, these teams felt they could not make the most of Essence due to their lack of knowledge about practices:

> *A team of beginner developers such as ourselves might get locked up in the [practice] cards already made, resulting in using methods that is ineffective for us since we wouldn't make up any new techniques that isn't "available". We think that with a little more experienced team that hasn't made their own method yet, this would be extremely helpful.* (Report 013)

Not all teams considered this to be a negative situation, however. Some teams felt that the way Essence encouraged them to experiment with new practices and to learn by working as a team was helpful, even though they initially did not have a clear idea of what practices might work for their team. Essence, they felt, challenged them to actively think about what they were doing and why, and even though it did not provide direct answers to those questions, it facilitated learning in a positive manner. Thus, the general sentiment among the groups was that Essence, as well as the practice library related to it, had been very useful for them as inexperienced developers. As a concluding remark, it is worth noting that while not all of the teams comprised of individuals with little or no past experience with practical SE work, the resounding majority of them nonetheless did, being comprised of second year SE students. This was also evident in the way the teams actively reflected on their own inexperience in various ways in their reports.

### 3.3 Theme 3: Way of Working and the Method Prison

One of the most discussed positive aspects of Essence perceived by the teams was its method-agnostic approach. The ability to freely choose between methods and practices was considered both new and highly positive, letting them, in the words of Jacobson et al. (2017), escape the "method prison":

> *Our team really liked the freedom SEMAT gives you in defining the way you develop something and how you can customize it, choose the practices you want and not be forced to use practices you don't want to use* (Report 036)

> *There were many positives of applying the kernel to our project, like choosing what we wanted to implement in our regular work day allowed us to use only what we wanted and thought we could benefit from. This level of freedom created a higher level of productivity than for example Scrum, where we are forced to use all aspects of the framework that do not necessarily benefit us. Not being forced to do things that we feel would slow us down and not benefit us really made us appreciate the SEMAT Essence Kernel* (Report 071)

As many of the students in the course had previously taken a course on Scrum, many of the reports consequently also included reflections related to Scrum. These teams discussed how they had initially started using Scrum or ScrumBut but had then begun to reflect on what they were doing and why, resulting in them refining their own way of working by using Essence. Used in conjunction with the practice card library, Essence provided them with new alternative practices to utilize. This resulted in the teams experimenting with different practices. On a more general level, they felt that the method-agnostic approach of Essence prepared them for different ways of working in the future.

Additionally, the teams reported positive experiences with actively reflecting on their way of working. Aside from initially tailoring a method for themselves, some of the teams reported having found Essence useful in facilitating the idea of continuously improving their work processes based on their experiences. Furthermore, some teams also noted that Essence had made it easier to communicate their way of working to the team as well as to discuss it within the team:

> *This overview of all practices really benefited us when we put together our way of working and made it easy to visualize our workflow. Whenever a team member was unhappy with any aspect of our work methodology we reviewed the cards and added or removed any if needed.* (Report 060)

Finally, the teams discussed having learned much about new methods and practices simply by browsing through the practice cards available in the Ivar Jacobson practice library. This serves to underline the importance of tools related to adopting Essence. In this case, the practice cards helped teams of inexperienced developers tailor methods using Essence despite not having any previous experience with different SE practices.

### 3.4 Theme 4: Progress Control

The Essence kernel provides a framework upon which to build a project-specific tool. However, even without any modifications, the kernel already serves as a basic progress control tool. This was also reflected in the reports. Most teams that had properly utilized the kernel had had positive experiences using Essence to manage and track progress:

> *Selecting and using the alpha state cards that were relevant to our circumstances to assess our progress proved extremely effective. When we used them for the first time we were surprised to learn that we had not made as much progress as we thought. The cards were useful in seeing where we wanted to be in terms of progress in the different alphas, and thus facilitated the process of fixing our impediments.* (Report 005)

> *The team then agreed to purchase a cork board and print out the Alpha State Cards in order to quickly and easily get an overview over the team's overall progress. This proved valuable, as none of the team members had partaken in any projects of this scale previously. The clear visualization the cards provided gave a much clearer picture of the project's progression overall than what the team found orally.* (Report 055)

Although Essence did clearly facilitate the idea of tailoring methods and choosing the methods that work best, this may not always be preferable. If the alternative to being locked in a "method prison" is the use of ineffective ad hoc methods, following an established method by the book may well be the more effective option. However, the teams felt that Essence helped them *formalize* their way of working aside from also facilitating the idea of tailoring it to suit their context-specific needs.

In relation to the inexperience of the teams discussed in a preceding sub-section, many of the teams felt that the Essence kernel provided a good overview of a software engineering endeavor *especially* because they had little experience with SE project work. Even though not all teams that utilized the kernel extended it, they nonetheless felt the Essence kernel in its base form was already useful in tracking their progress – except for one. One of the teams felt that they had had a solid understanding of the state of their project prior to using Essence and that "it didn't help us anything to convert it into cards and more complicated sentences" (Report 059). This is not surprising as tools are just that: tools. Similarly, though formal methods and practices are typically preferred, it is quite possible to carry out SE endeavors using ad hoc methods, as e.g. a notable number of software startups chooses to do [17].

### 3.5 Summary of Findings

Having discussed the results through the themes present in the data set, we now turn back to our formal research problem. Below, we provide summarizing answers for the two research questions posed in the introduction before going into more detail:

**RQ1:** Do bachelor level students find Essence useful?
**Results:** Essence was considered useful by the students, for varying reasons

**RQ2:** What are the challenges in adopting Essence, specifically for inexperienced software developers, and what could be done to make its adoption easier?
**Results:** The largest challenge in adopting Essence was the lack of good tutorial resources, which consequently could be addressed by creating better such resources.

Though the student teams nearly universally considered Essence useful, there were differences between the teams in terms of *why* they considered it useful, largely based on the extent to which they had utilized it. Essence was considered useful for (1) teaching new methods and practices, (2) teaching a method-agnostic approach to SE, (3) helping the team properly structure their way of working, and (4) providing a useful framework for managing an SE project, depending on the degree of its utilization among each team. Few teams had anything negative to say about the specification itself, with most of the negative feedback relating to difficulties in adopting Essence.

Indeed, though Essence was considered useful by the teams, it was nonetheless evidently difficult for them to adopt. Many teams, even those that did utilize it the most, considered it to have been difficult to initially learn. The reports that discussed the reasons behind its perceived difficult adoption all cited the lack of good tutorial resources as the main problem. The teams felt that the resources they could find online were either hundreds of pages long or did simply not describe Essence simply enough for beginners. This resulted in some teams opting to focus their efforts elsewhere by e.g. focusing on learning to program and use programming tools, leaving Essence for later.

Having discussed our findings in relation to our research questions, we present a further, visual summary of how the themes discussed earlier in this section are interlinked (Fig. 3). It is organized in a manner similar to how Giardino et al. [6] summarized their findings and depicts the adoption of Essence among students as a process. The student teams, as developers, were inexperienced. This inexperience resulted in a lack of resources as they had to divide their resources between e.g. learning to program, learning to use the programming tools, and learning Essence. In this situation, Essence often took on a lower priority, consequently becoming more difficult for the teams to learn. However, once the teams began to understand and utilize Essence, they began to work more systematically. All teams utilized Essence and the practice cards to work in a more systematic fashion, and many, but not all, teams grasped the kernel and began to use it as a progress control tool. For the teams that understood how to fully utilize Essence, its use ultimately resulted in an escape from the so-called method prison [10]. These teams actively reflected on their way of working and saw Essence also as a tool to facilitate learning in order to (attempt to) work in an efficient fashion in any given context in the future.

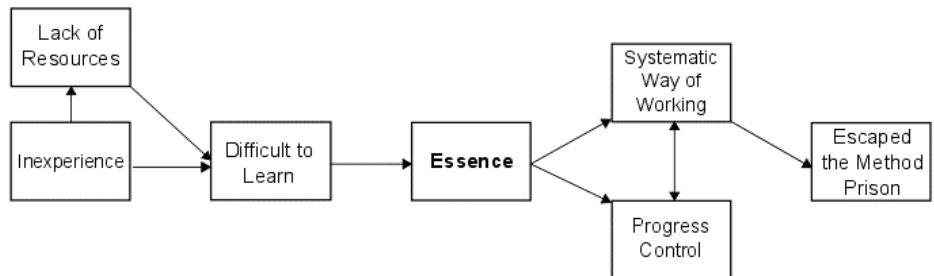

**Figure 3**. Adoption process of Essence among SE students

Based on our findings, we therefore argue that SE students find Essence useful for multiple reasons. Furthermore, we confirm that Essence is considered difficult to learn, and our data suggests that the largest challenges in adopting Essence currently stem from a lack of tutorials and guides aimed at beginners. The current resources available online were considered too lengthy or advanced to be of use for new users of Essence.

## 4    Discussion

As extant literature has suggested [8], our findings confirm that Essence is indeed considered difficult and resource-intensive to adopt. However, our findings indicate that stems from a lack of good tutorial resources as opposed to Essence being difficult to use as such. The current manuals and other resources available were considered by the student teams to be too complex for beginners. Thus, the most direct solution to this issue would simply be the creation of better tutorial resources specifically aimed at new users of Essence.

As a solution to making Essence easier to adopt, [8] suggested the development of tools that could be used to make the practical use of Essence easier. This was not confirmed by our findings as none of the teams voiced explicit wishes for more tools to help utilized Essence. However, given that the practice card library, an external tool as well, was very positively received among the teams, it is likely that further tooling would also make Essence either easier to adopt and possibly more useful.

In terms of the usefulness of Essence for bachelor level students, our data indicates that Essence was indeed considered useful by the resounding majority of the project teams we studied. Less than ten teams out of 102 reported having found the use of Essence an outright negative and useless experience. In this light, we argue that Essence is useful for bachelor level students. More specifically, it was found useful in terms of (1) teaching new methods and practices, (2) teaching a method-agnostic approach to SE, (3) helping the team properly structure their way of working, and (4) providing a useful framework for managing an SE project.

From the point of view of SE education in universities, Essence is interesting as, based on our experiences, it can potentially provide a common ground for SE education through its method-agnostic nature. Such common ground is currently missing. We have showed that it can simultaneously teach students SE progress control as well as practical SE work. It also prepares SE students for working with different methods and practices out on the field. Essence could therefore be used to provide students with a higher-level understanding of the way SE work is structured. Essence can serve as a basis upon which SE students can build a general understanding of different SE methods as opposed to learning about single methods one at a time.

Learning to construct a method out of practices is an important learning goal for software engineering education. Based on our observations during the course, it was noted that some teams also learned to include so called anti-patterns or bad practices explicitly in their process description. This is a novel thought and should be further elaborated in future studies. By labeling a practice as a bad-practice, the team in question explicitly communicated about their improvement needs. Manual testing is an example of such practice as it indicates lack of automated test suite, which slows down the development and is thus not a sustainable solution.

Additionally, in terms of generalizing our findings, we suggest that our findings could also be interesting for future research from the point of software startups. SE students, like startup practitioners [3, 12], are often more inexperienced developers, and it is also not uncommon for university students to participate in software startups during their studies. Most software startups fail [7] for various reasons, and Kon et al. [12] posited that specifically younger, more inexperienced startup practitioners are considered more prone to failure among investors. Software startups face various challenges across their life cycles [22], including challenges with "building product", "staying focused & disciplined", and "over capacity/too much to do", which Essence could potentially be used to aid in solving. Finally, it has been established that software startups, like mature organizations, should concern themselves with structuring their work processes [19], which is something we found Essence to be useful for among SE students. Relating these past studies to our findings here, we suggest that future studies could investigate Essence from the point of view of software startups. Our findings, however, do not offer direct support to this link between these two contexts. In possibly pursuing this line of research, it could be useful to also evaluate the suitability of the Essence kernel in the context of software startups, as software startups have been shown to develop software in different ways than mature organizations [10], and their business aspect is linked with their SE process in a unique fashion.

Finally, while we have studied perceived difficulties in adopting Essence in the context of SE students, future studies may wish to study impediments to its adoption among practitioner organizations. As Essence has yet to see widespread practitioner adoption [21], the reasons behind this situation are worth investigating. Similarly, it is likely that more experienced practitioners find Essence useful or not useful for different reasons than the SE students studied in this paper.

### 4.1 Limitations of the Study

The primary limitations of the study are associated with the data collected during it. In collecting the data, we chose to rely on self-reported use of Essence over observation and regular check-ups. From this results that the validity of the reported utilization of Essence among the teams cannot be directly confirmed. However, the student teams seldom failed to report problems in utilizing Essence, with most teams that failed to utilize Essence fully reporting so themselves. In other cases, it was also largely possible to determine whether a team had understood the specification or not based on the way they reported on its utilization. We thus argue that this does not present a major threat to the validity of our data in such a large data set (102 teams).

Additionally, while the use of students as subjects for scientific studies is a long-standing topic of discussion across disciplines, including SE, the aim of this study was to study Essence specifically in relation to SE students and education. The use of students as subjects in this context is therefore not an issue.

## 5 Conclusions

In this paper, we have studied the Essence Theory of Software Engineering in a large-scale bachelor level course through experience reports. We introduced Essence to 102 project teams in a project-based SE course at a Norwegian university and observed its use during the projects. Based on 102 project reports discussing, among other things, the Essence use experiences of project teams of 4-5 individuals, we described the barriers of adoption of Essence and its usefulness for SE students.

We discovered that while Essence was considered difficult to learn by the teams, these difficulties largely stemmed from the lack of good tutorial resources. Some teams failed to fully utilize Essence, forgoing its progress control aspect partially or entirely, primarily due to its difficult adoption. There is thus a clear need for better introductory guides to Essence that are specifically designed for new users.

Past its difficult adoption, Essence was nonetheless nearly universally considered useful by the project teams. Even the teams that had not fully utilized Essence considered the method-agnostic approach and the practice cards to have been useful for planning out and formalizing their way of working during their projects. Additionally, the teams that had grasped the Essence kernel (except for two teams) also reported Essence having been useful in tracking progress during their projects. They felt that Essence gave them a good general understanding of SE project work through the alphas and that the alpha states helped them keep track of progress on their endeavor.

We therefore argue in favour of using Essence in SE education. By helping SE students gain a better understanding of SE project work and by preparing them for future adoption of various practices and methods, Essence can help tackle gaps [2, 13] between SE education and practice. To summarize our findings:

(1) Essence can teach students new methods and practices by encouraging them to study them in order to tailor their own methods using Essence
(2) Essence encourages students to adjust their way of working based on the SE context at hand as opposed to following existing methods by the book
(3) Essence helps students structure their way of working in a practical setting
(4) Better tutorial resources for Essence are needed to make it easier to adopt

## References

1. Abrahamsson, P., Salo, O., Ronkainen, J., and Warsta, J.: Agile Software Development Methods: Review and Analysis. Otamedia: VTT Publications, 478 (2002).
2. Almi, N. E. A. M., Rahman, N. A., Purusothaman, D., and Sulaiman, S.: Software engineering education: The gap between industry's requirements and graduates' readiness. In: Computers & Informatics (ISCI), 2011 IEEE Symposium on (2011).
3. Crowne, M.: Why software startups fail and what to do about it – Evolution of software product development in startup companies. In: Proceedings International Engineering Management Conference (IEMC), 338-343 (2002).
4. Davis, F. D. Jr.: A Technology Acceptance Model for Empirically Testing New End-User Information Systems: Theory and Results. Massachusetts Institute of Technology (1985).
5. Ghanbari, H.: Investigating the causal mechanisms underlying the customization of software development methods. Uni. of Jyväskylä: Jyväskylä Studies in Computing, 258 (2017).
6. Giardino, C., Paternoster, N., Unterkalmsteiner, M., Gorschek, T., and Abrahamsson, P.: "Software Development in Startup Companies: The Greenfield Startup Model". IEEE Transactions on Software Engineering, 42(6), pp. 585-604 (2016).
7. Giardino, C., Wang, X., and Abrahamsson, P.: Why Early-Stage Software Startups Fail: A Behavioral Framework. In International Conference of Software Business, pp. 27-41. Springer, Cham (2014).
8. Graziotin, D., and Abrahamsson, P.: A Web-based modeling tool for the SEMAT Essence theory of Software Engineering. Journal of Open Research Software, 1 (2013).
9. Jacobson, I., Stimson, R., and Hastie, S.: Escaping Method Prison. https://www.infoq.com/articles/escape-method-prson last accessed 15 May 2018 (2017)
10. Jacobson, I., Ng, P., McMahon, P. E., Spence, I., and Lidman, S.: The Essence of Software Engineering: The SEMAT Kernel. ACMQueue, 10, pp. 40-52 (2012).
11. Kemell, K. O., Risku, J., Evensen, A., Dahl, A. M., Grytten, L., Jedryszek, A., Rostrup, P., Nguyen-Duc, A., and Abrahamsson, P.: Gamifying the Escape from the Engineering Method Prison - An Innovative Board Game to Teach the Essence Theory to Future Project Managers and Software Engineers. [to be published in the proceedings of ICE2018] (2018)
12. Kon, F., Cukier, D., Melo, C., Hazzan, O., and Yuklea, H.: A Panorama of the Israeli Software Startup Ecosystem. SSRN: https://ssrn.com/abstract=2441157 (2014).
13. Lethbridge, T. C., Díaz-Herrera, J., LeBlanc Jr., R. J., and Thompson, J. B.: Improving software practice through education: Challenges and future trends. Proceedings: FOSE '07 Future of Software Engineering (2007).
14. Object Management Group: Essence – Kernel and Language for Software Engineering Methods. Version 1.1. http://semat.org/essence-1.1, last accessed 2018/05/28.
15. Park, J. S., McMahon, P. E., and Myburgh, B.: Scrum Powered by Essence. ACM SIGSOFT Software Engineering Notes, 41(1), pp. 1-8 (2016).
16. Parnin, C., Helms, E., Atlee, C., Boughton, H., Ghattas, M., Glover, A., Holman, J., Micco, J., Murphy, B., Savor, T., Stumm, M., Whitaker, S., and Williams, L.: The Top 10 Adages in Continuous Deployment. IEEE Software, 34(4), 86-95 (2017).
17. Paternoster, N., Giardino, C., Unterkalmsteiner, M., Gorschek, T. and Abrahamsson, P.: Software development in startup companies: A systematic mapping study. Information and Software Technology, 56, pp. 1200-1218 (2014).
18. Pieper, J.: Discovering the Essence of Software Engineering – An Integrated Game-Based Approach based on the SEMAT Essence Specification. In Proceedings of the 2015 IEEE Global Engineering Education Conference (EDUCON), pp. 939–947 (2015).
19. Ries, E.: The Lean Startups: How Today's Entrepreneurs Use Continuous Innovation to Create Radically Successful Businesses. New York: Crown Books (2011).
20. SEMAT: SEMAT and Essence – What is it and why should you care? http://semat.org/what-is-it-and-why-should-you-care-, last accessed 2018/05/20.
21. SEMAT: Great pick up of Semat. http://semat.org/news/-/asset_publisher/eaHEtyeuE9wP/content/great-pick-up-of-semat, last accessed 2018/05/13.
22. Wang, X., Edison, H., Bajwa, S. S., Giardino, C., and Abrahamsson, P.: Key Challenges in Software Startups Across Life Cycle Stages. In: Sharp H., Hall T. (eds) Agile Processes, in Software Engineering, and Extreme Programming. XP 2016. Lecture Notes in Business Information Processing, vol 251. Springer, Cham (2016)